\documentstyle[11pt,moriond,epsfig]{article}

\def\be{\begin{equation}}
\def\ee{\end{equation}}
\def\bea{\begin{eqnarray}}
\def\eea{\end{eqnarray}}

\newcommand{\W}{\mbox{$W$}}

\newcommand{\qsq}{\mbox{$Q^2$}}

\newcommand{\rh}{\mbox{$\rho$}}

\newcommand{\ph}{\mbox{$\phi$}}
\newcommand{\om}{\mbox{$\omega$}}

\newcommand{\jpsi}{\mbox{$J/\psi$}}

\newcommand{\rzqzz}{\mbox{$r_{00}^{04}$}}
\newcommand{\rzquz}{\mbox{$r_{10}^{04}$}}
\newcommand{\rzqumu}{\mbox{$r_{1-1}^{04}$}}
\newcommand{\ruuu}{\mbox{$r_{11}^{1}$}}
\newcommand{\ruzz}{\mbox{$r_{00}^{1}$}}
\newcommand{\ruuz}{\mbox{$r_{10}^{1}$}}
\newcommand{\ruumu}{\mbox{$r_{1-1}^{1}$}}
\newcommand{\rduz}{\mbox{$r_{10}^{2}$}}
\newcommand{\rdumu}{\mbox{$r_{1-1}^{2}$}}
\newcommand{\rcuu}{\mbox{$r_{11}^{5}$}}
\newcommand{\rczz}{\mbox{$r_{00}^{5}$}}
\newcommand{\rcuz}{\mbox{$r_{10}^{5}$}}
\newcommand{\rcumu}{\mbox{$r_{1-1}^{5}$}}
\newcommand{\rsuz}{\mbox{$r_{10}^{6}$}}
\newcommand{\rsumu}{\mbox{$r_{1-1}^{6}$}}

\newcommand{\cosths}{\mbox{$\cos\theta$}}

\newcommand{\gevsq}{\mbox{${\rm GeV}^2$}}

\newcommand{\bce}{\begin{center}}
\newcommand{\ece}{\end{center}}
\newcommand{\beq}{\begin{equation}}
\newcommand{\eeq}{\end{equation}}
\def\lsim{\mathrel{\rlap{\lower4pt\hbox{\hskip1pt$\sim$}}
    \raise1pt\hbox{$<$}}}         
\def\gsim{\mathrel{\rlap{\lower4pt\hbox{\hskip1pt$\sim$}}
    \raise1pt\hbox{$>$}}}         
\def\ar#1#2#3   {{\em Ann. Rev. Nucl. Part. Sci.} {\bf#1} (#2) #3}
\def\err#1#2#3  {{\it Erratum} {\bf#1} (#2) #3}
\def\ib#1#2#3   {{\it ibid.} {\bf#1} (#2) #3}
\def\ijmp#1#2#3 {{\em Int. J. Mod. Phys.} {\bf#1} (#2) #3}
\def\jetp#1#2#3 {{\em JETP Lett.} {\bf#1} (#2) #3}
\def\mpl#1#2#3  {{\em Mod. Phys. Lett.} {\bf#1} (#2) #3}
\def\nim#1#2#3  {{\em Nucl. Instr. Meth.} {\bf#1} (#2) #3}
\def\nc#1#2#3   {{\em Nuovo Cim.} {\bf#1} (#2) #3}
\def\np#1#2#3   {{\em Nucl. Phys.} {\bf#1} (#2) #3}
\def\pl#1#2#3   {{\em Phys. Lett.} {\bf#1} (#2) #3}
\def\prep#1#2#3 {{\em Phys. Rep.} {\bf#1} (#2) #3}
\def\prev#1#2#3 {{\em Phys. Rev.} {\bf#1} (#2) #3}
\def\prl#1#2#3  {{\em Phys. Rev. Lett.} {\bf#1} (#2) #3}
\def\ptp#1#2#3  {{\em Prog. Th. Phys.} {\bf#1} (#2) #3}
\def\rmp#1#2#3  {{\em Rev. Mod. Phys.} {\bf#1} (#2) #3}
\def\rpp#1#2#3  {{\em Rep. Prog. Phys.} {\bf#1} (#2) #3}
\def\sjnp#1#2#3 {{\em Sov. J. Nucl. Phys.} {\bf#1} (#2) #3}
\def\spj#1#2#3  {{\em Sov. Phys. JEPT} {\bf#1} (#2) #3}
\def\zp#1#2#3   {{\em Zeit. Phys.} {\bf#1} (#2) #3}
\def\eur#1#2#3  {{\em Eur. Phys. Jour.} {\bf#1} (#2) #3}
\begin{document}
\vspace*{4cm}
\title{HARD DIFFRACTION IN VECTOR MESON PRODUCTION AT HERA \\
\vspace*{0.4cm} On behalf of the H1 and ZEUS Collaborations}

\author{B. CLERBAUX }

\address{Universit\'e Libre de Bruxelles - CP 230, B-1050 Brussels, Belgium
\\e-mail: bclerbau@hep.iihe.ac.be}

\maketitle\abstracts{
Results on elastic vector meson production at HERA are presented in
the framework of perturbative QCD. The energy dependence of the
cross section for \jpsi\ photoproduction and for \rh\ electroproduction
is studied. A full polarisation analysis, including the measurement of
the 15 elements of the spin density matrix and of $R=\sigma_L / \sigma_T$,
is presented for \rh\ electroproduction. Finally  heavier vector
mesons ($\psi'$ and $\Upsilon$) production is discussed.}

\section{Introduction}

\begin{figure}[tb]
\begin{center}
  \setlength{\unitlength}{1cm}
  \begin{picture}(16.0,6.0)
   \put(-0.4,0.15){\epsfig{file=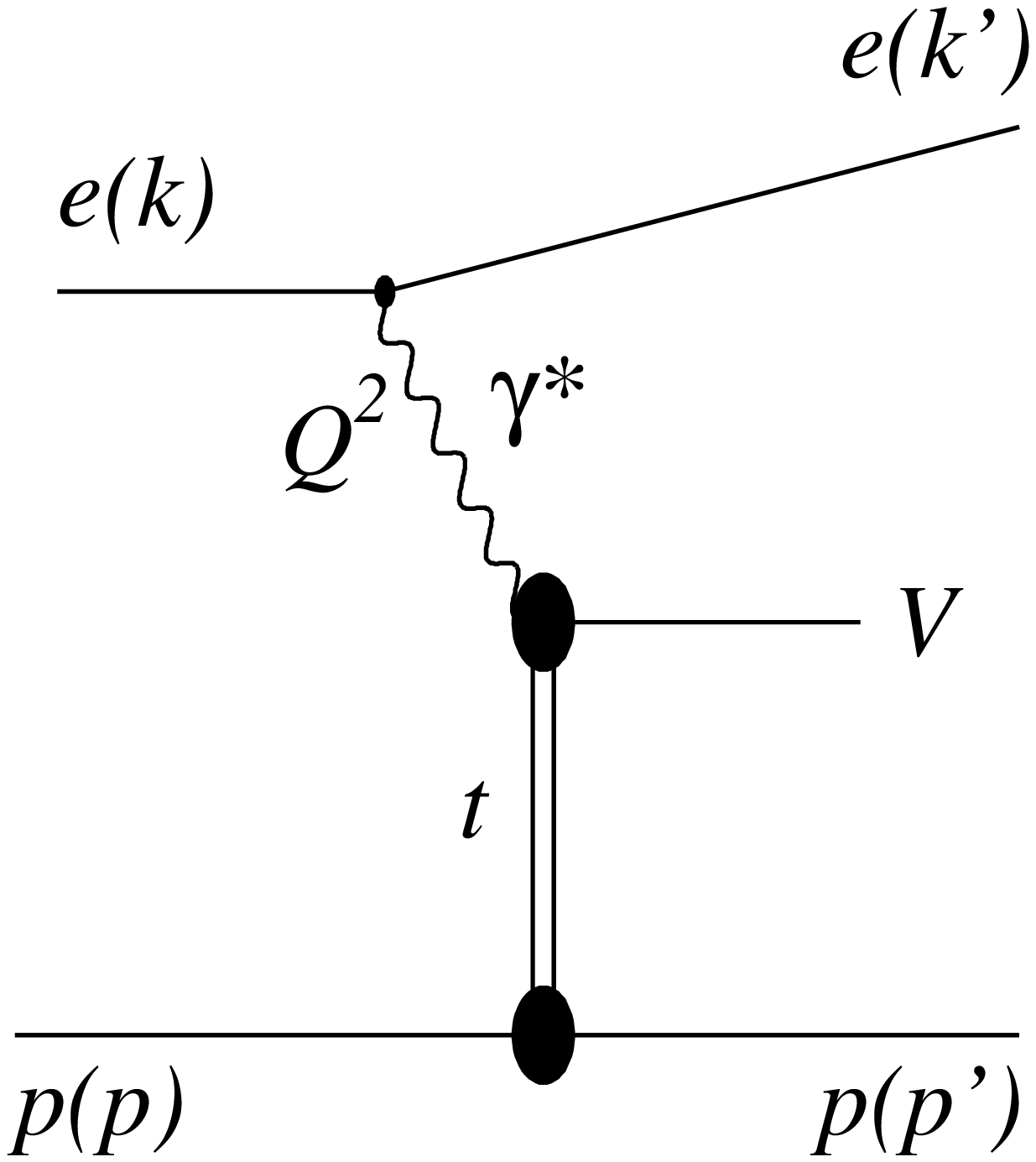,%
      width=5cm,height=5.5cm}}
   \put(4.2,0.0){\epsfig{file=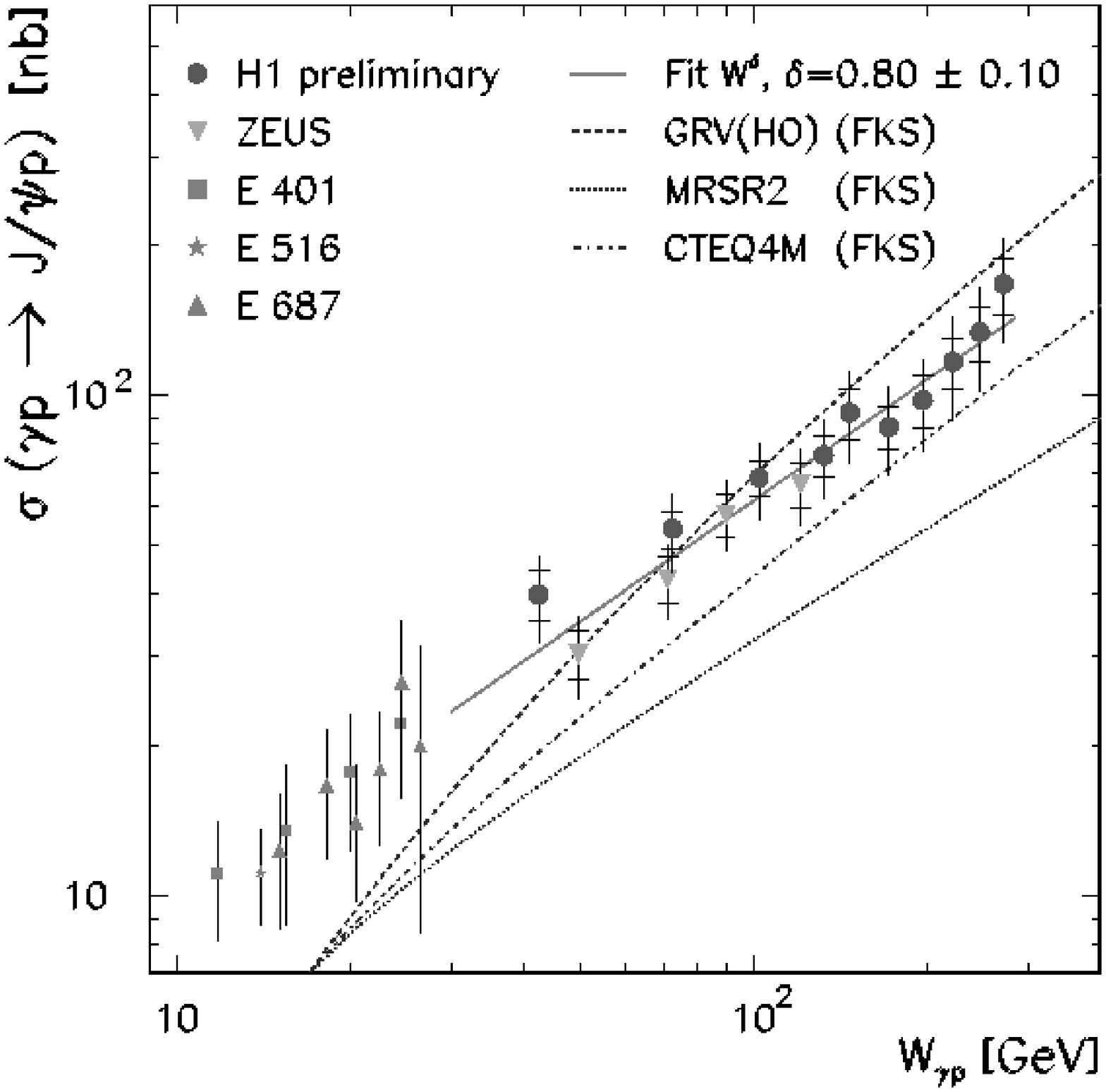,%
      bbllx=0pt,bblly=0pt,bburx=590pt,bbury=580pt,%
      width=5.5cm,height=5.5cm}}
   \put(10.0,0.0){\epsfig{file=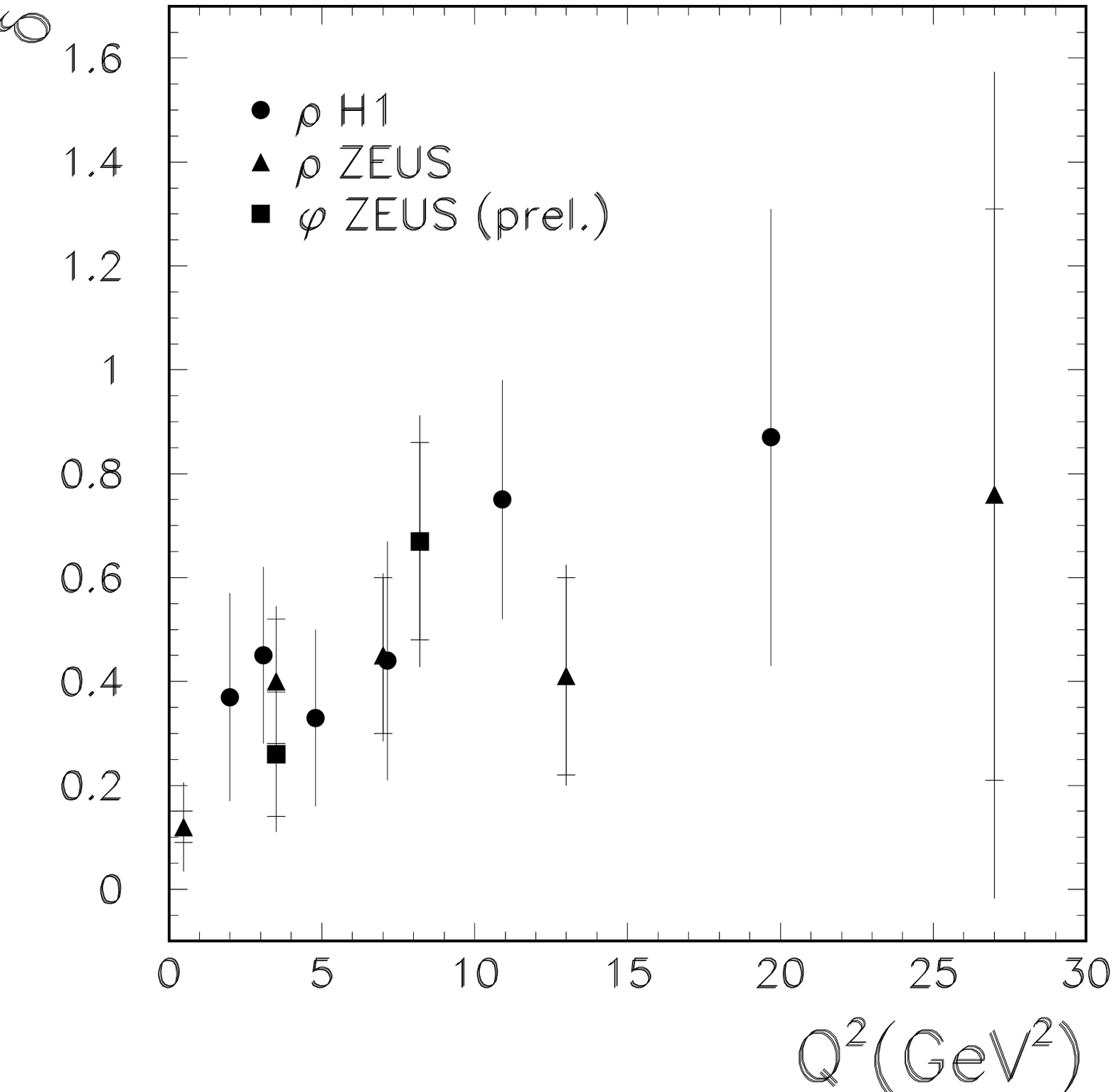,
         width=6cm,height=6cm}}
   \put(1.9,0.0){(a)}
   \put(7.1,0.0){(b)}
   \put(12.7,0.0){(c)}
  \end{picture}
  \caption{
  a) Diagram for the reaction $e+p \rightarrow e+p+V$; 
  b) Cross section $\sigma (\gamma p \rightarrow \jpsi p)$ for \jpsi\ meson
     photoproduction 
     as a function of \W. The dotted lines represent predictions of 
     a perturbative QCD model~\protect\cite{fks} using different
     gluon density parametrisations and the solid line represents a fit to the
     data using the parametrisation $\sigma \propto W^\delta$, with $\delta$ = 0.80 $\pm$ 0.10;  
  c) \qsq\ dependence of the $\delta$ parameter for \rh\ and \ph\ electroproduction.}
  \label{fig:cross}
\end{center}
\end{figure}

Elastic electroproduction of vector mesons,
$ e+p \rightarrow e+p+V$, is studied at HERA~\cite{jpsi_h1,rho_h1,gp_psip_h1,rho_zeus,upsilon_zeus} in a
wide kinematical range in \qsq\ (the photon virtuality), \W\
(the $\gamma^{(*)} p$ centre of mass energy), $t$ (the 
square of the 4-momentum exchanged at the proton vertex)
and $m_V$ (the vector meson mass) (see Fig.~\ref{fig:cross}a).
The vector mesons
$V = \rho, \om, \ph, \jpsi, \psi'\ {\rm and}~\Upsilon$ 
have the same quantum numbers as the 
photon ($J^{PC} = 1^{--}$) and the $\gamma^{(*)} p$ interaction 
is mediated by the exchange of a colourless
object, the pomeron.
Pomeron exchange was studied in detail in the framework of the Regge model; 
interest is now to understand it in terms of partons in 
the framework of the QCD theory.
An important result of HERA studies is that, 
in contrast to the slow increase with the energy of the hadron-hadron cross sections
(``soft" behaviour), the total $\gamma^* p$ cross section has
a strong (``hard") energy dependence, which is attributed to a 
fast rise with energy of the gluon density in the proton. 
The QCD pomeron being described as a gluonic system, 
a ``hard" behaviour is thus also expected in vector meson production.

\section{Vector meson production in pQCD}

Quantitative predictions in perturbative QCD are possible when
a hard scale is present in the interaction. This scale
can be given by \qsq, $|t|$ or $m_q$, the quark mass.
Most models rely on the fact that, at high energy, 
in the proton rest frame, the photon fluctuates into a 
$q\bar{q}$ pair a long time before the interaction, and recombines 
into a vector meson (VM) a long time after the interaction.
The amplitude ${\cal M}$ then 
factories in three terms:
${\cal M} \propto \psi_{\lambda_{V}}^{V \ *}
 \  T_{\lambda_{V} \lambda_{\gamma}} \  \psi^\gamma_{\lambda_{\gamma}}$ 
where  $T_{\lambda_{V} \lambda_{\gamma}}$ are the interaction helicity amplitudes
($\lambda_\gamma$ and $\lambda_V$ being the helicities of the
photon and the VM respectively) and $\psi$ represents the wave functions.
In most models, the $q\bar{q}-p$ interaction is described by a 2 gluon
exchange. The cross section is then proportional to the square of the
gluon density in the proton:
%
$ \sigma_{\gamma p} \sim  \alpha_s^2(Q^2) / Q^6 \cdot \left
| xg(x, Q^2) \right| ^2  \label{eq:gluon} $.
The main uncertainties of the models come from the  
choice of the scale, of the gluon distribution
parametrisation, of the VM wave function (Fermi motion), and from the neglect of 
off-diagonal gluon distributions and of higher order corrections.

\section{Energy dependence of the cross sections}
At high energy, the total, elastic, and diffractive cross sections 
are dominated by pomeron exchange and 
the energy dependence of the cross sections
exhibits the ``soft" behaviour, parametrised as
$\sigma \propto W^{\delta}$, with $\delta = 0.22$.
The \rh, \om\ and \ph\ photoproduction cross sections
measured by the fixed target experiments and at HERA
present the same ``soft" dependence. 
However the \jpsi\ photoproduction cross section, where the mass of the
c quark provides a hard scale in the interaction, presents a much 
stronger energy dependence (``hard" behaviour)~\cite{jpsi_h1_prel}, 
in agreement with the rise of the  gluon density at low $x$ ($x$ $\simeq$ \qsq / $W^2$).
Figure~\ref{fig:cross}b presents the HERA measurement 
together with predictions of a 
perturbative QCD model~\cite{fks} using three 
parametrisations for the gluon density: GRVHO, MRSR2 and CTEQ4M. 
The full line represents a fit to the data using the parametrisation 
$\sigma \propto W^{\delta}$ with $\delta = 0.8 \pm 0.1$. This value of $\delta$ is
in contrast with the value $\delta$ = 0.22 obtained for the ``soft" energy dependence
of light vector mesons photoproduction. \\

Another way to look at hard behaviour is to study 
light vector meson production at high \qsq, \qsq\ giving here the scale.
Measurements of the cross section 
$\sigma ( \gamma^* p \rightarrow \rho p )$
show an indication for an increasingly stronger energy dependence when
\qsq\ increases~\cite{rho_h1,rho_zeus}. With the parametrisation 
$\sigma  \propto W^{\delta}$, we observe (see Fig.~\ref{fig:cross}c)
that at high \qsq\ the value of $\delta$ for the \rh\ meson production seems to reach
the one for \jpsi\ photoproduction.

\section{Polarisation studies}

\begin{figure}[hbt]
\begin{center}
\setlength{\unitlength}{1cm}
  \begin{picture}(20.0,11.0)
  \put(-1.3,0.0){\epsfig{file=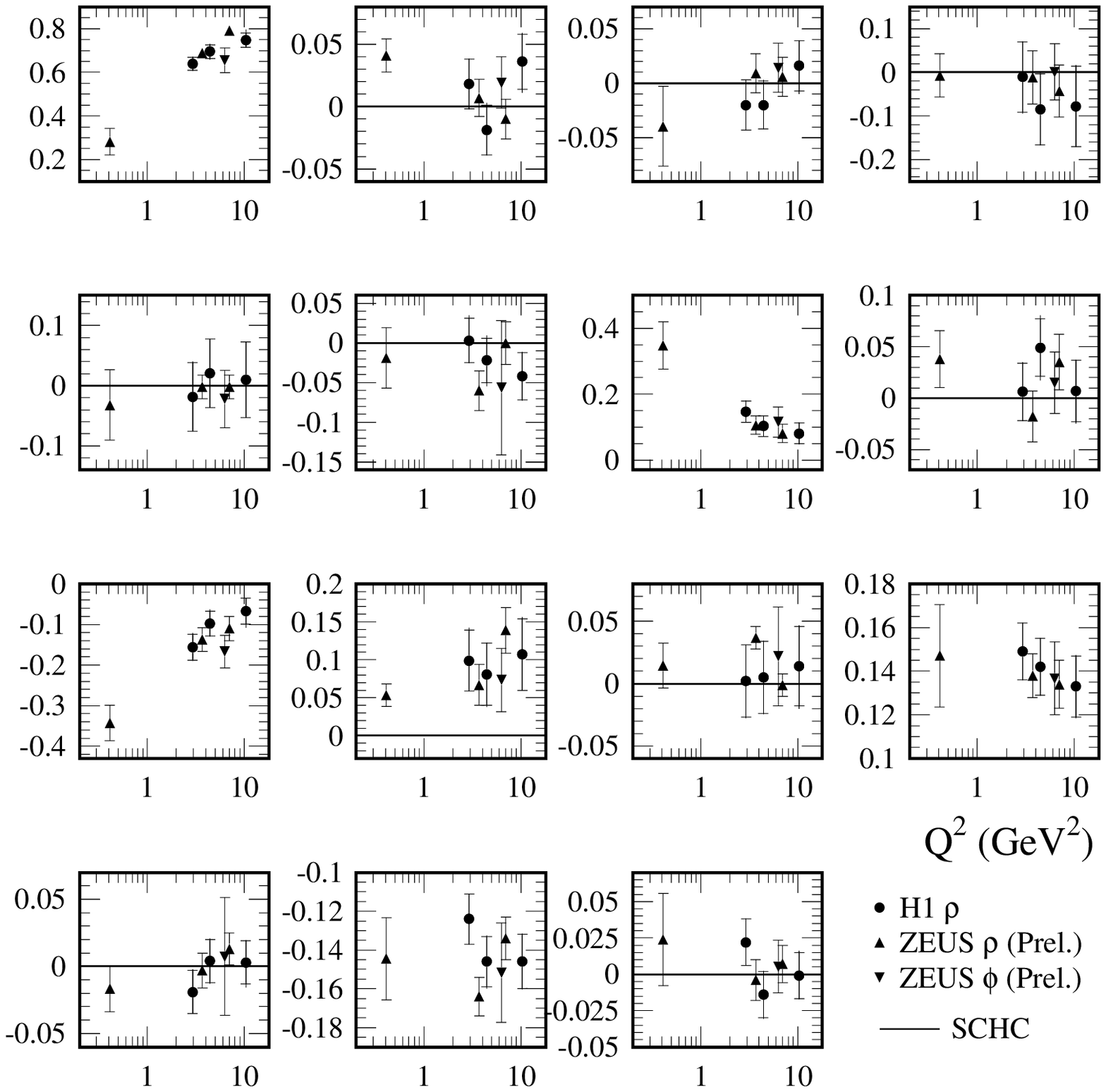,%
        width=11.5cm,height=11.5cm}}
  \put(0.9,10.55){\small \rzqzz}
  \put(3.0,10.55){\small {\rm Re} \rzquz}
  \put(5.5,10.55){\small \rzqumu}
  \put(7.9,10.55){\small \ruzz}
  \put(0.9,8.1){\small \ruuu}
  \put(3.0,8.1){\small {\rm Re} \ruuz}
  \put(5.5,8.1){\small \ruumu}
  \put(7.7,8.1){\small {\rm Im} \rduz}
  \put(0.65,5.6){\small {\rm Im} \rdumu}
  \put(3.2,5.6){\small \rczz}
  \put(5.55,5.6){\small \rcuu}
  \put(7.7,5.6){\small {\rm Re} \rcuz}
  \put(0.8,3.1){\small \rcumu}
  \put(3.0,3.1){\small {\rm Im} \rsuz}
  \put(5.24,3.1){\small {\rm Im} \rsumu}
  \put(9.5,3.0){\epsfig{file=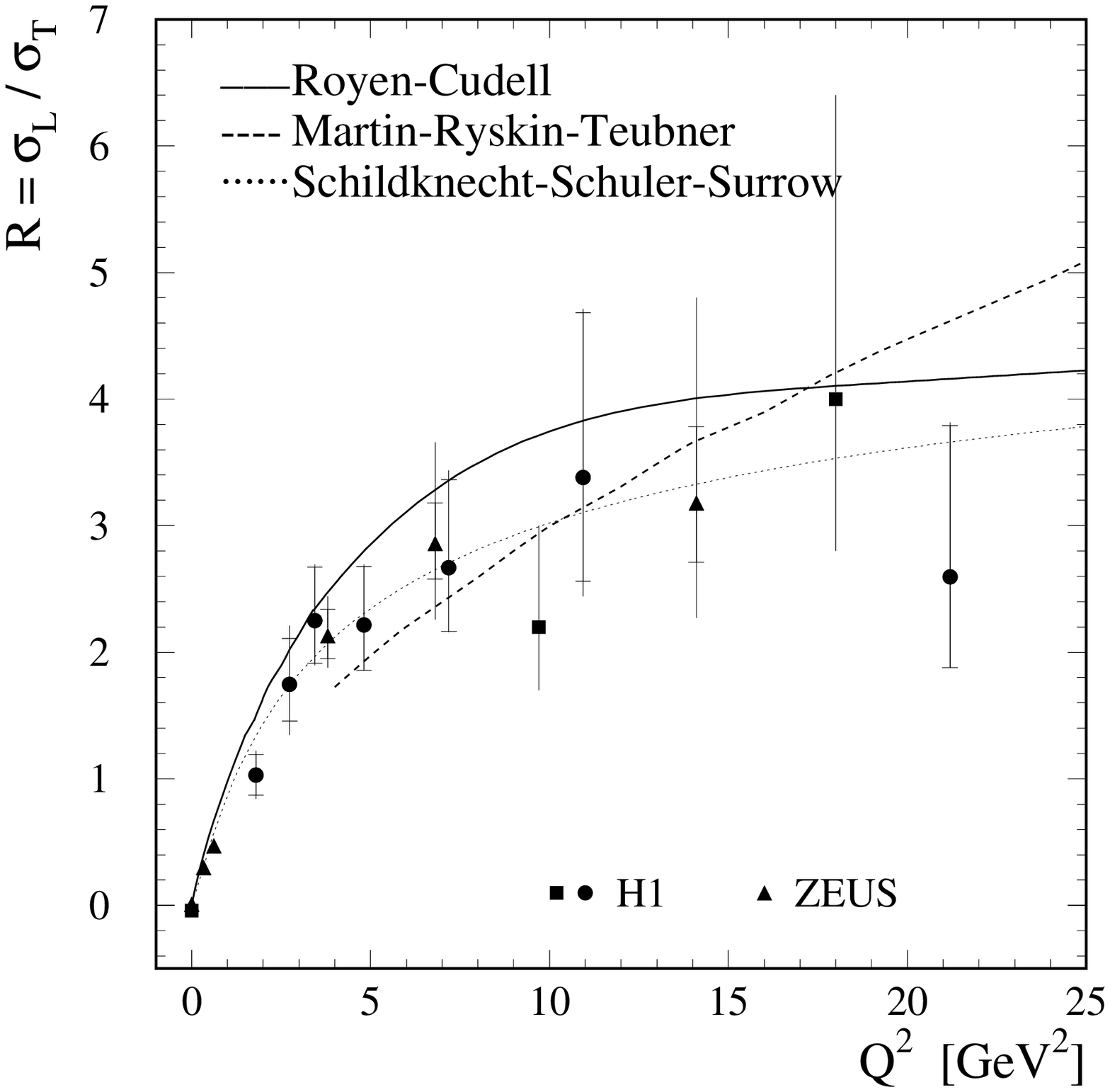,%
        width=7.0cm,height=7.0cm}}
  \put(10.0,2.0){$\leftarrow$ (a)}
  \put(13.0,2.5){(b)}
  \end{picture}
  \vspace*{-1.5cm}
  \caption{
  a) Measurements of the 15 elements of the \rh\ and \ph\ spin density 
      matrix as a function of \qsq. 
     The lines represent the zero predictions in case of SCHC;
  b) \qsq\ dependence of $R=\sigma_L / \sigma_T$,
     the lines being predictions of three models (see text).}
  \label{fig:pola}
\end{center}
\end{figure}

Full helicity studies have been performed for the \rh\ and \ph\ meson
electroproduction~\cite{rho_h1,hel_zeus_prel}. In the helicity frame, three angles are 
used: the polar ($\theta$) and azimuthal ($\varphi$) angles of the positive track
in the vector meson centre of mass system (cms), and the 
$\Phi$ angle between the electron scattering plane and the 
vector meson production plane, in the hadronic cms. The decay angular 
distribution $W(\cosths, \varphi, \Phi)$ is a function of 
15 matrix elements $r^{\alpha}_{ij}$
and $r^{\alpha \beta}_{ij}$, which are related to the 
helicity amplitudes $T_{\lambda_{V} \lambda_{\gamma}}$. The study of 
the angular distribution thus allows extracting information on the
helicity states of the exchanged virtual photon and of the vector 
meson in the final state. Figure~\ref{fig:pola}a presents the measurement
of the 15 matrix elements as a function of \qsq. In case of s-channel 
helicity conservation (SCHC), the helicity of the vector meson is the same
as that of the photon ($T_{01} = T_{10} = T_{1-1} = T_{-11} = 0$),
and 10 of the matrix elements vanish (lines in Fig.~\ref{fig:pola}a).
It is the case for 9 of them but not for the \rczz\ parameter which is found
to be significantly different from zero. The ratio of helicity flip to non helicity flip 
amplitudes is hence estimated to be $8.0 \pm 3.0$ \%. 
The ratio of the longitudinal to the transverse cross 
section, $R = \sigma_L / \sigma_T$, can be extracted using the \rzqzz\ matrix
element. $R$ is observed to 
increase with \qsq, and to reach the value
$R$ = 3 - 4 for \qsq\ $\simeq$ 20 \gevsq\ (see Fig.~\ref{fig:pola}b). 
The following hierarchy between the helicity amplitudes, observed in the data: 
$|T_{00}| > |T_{11}| > |T_{01}| > |T_{10}|, |T_{1-1}|$, is in agreement 
with perturbative QCD calculations performed by Ivanov and Kirschner~\cite{ivanov}.
The \qsq\ dependence of the ratio $R$ is well described by the perturbative QCD models of
Royen and Cudell~\cite{royen}, and of Martin, Ryskin and 
Teubner~\cite{mrt} and by the model of Schildknect, Schuler and Surrow~\cite{sss} 
based on generalised vector dominance model (GVDM).

\section{\boldmath $\psi'$ and $\Upsilon$ production}

Signals for $\psi'$ and $\Upsilon$ production have been 
observed recently at HERA~\cite{psip,upsilon_zeus,upsilon_h1}.
The ratio of $\psi'/ \psi$ production cross sections is measured for four 
different \qsq\ values
ranging from 0 to 25 \gevsq\ and an indication for a rise of the ratio as
a function of \qsq\ is observed. This can be understood as due to the 
``scanning'' of the VM wave function as \qsq\ varies. Indeed, in photoproduction,
the Compton wave length of the photon is comparable to the radius of the
$\psi$(2S) wave function. But the latter has a node which
induces approximately cancelling contributions to the production amplitude, implying that
the photoproduction of $\psi(2S)$ mesons is small. As \qsq\ 
increases, the transverse size of the $q \bar q$ pair decreases,
thus avoiding the cancellation effect.  \\

The value of 
$\sigma (\gamma p \rightarrow \Upsilon p) * {\rm BR} (\Upsilon 
\rightarrow \mu^+ \mu^-$) for \qsq\ $\simeq$ 0 is measured
to be, respectively, $16.0 \pm 8.5$ and $13.0 \pm 6.6$ pb by the H1 and ZEUS collaborations;
due to limited statistics the data cannot distinguish between 1S, 2S
and 3S states of the $\Upsilon$ meson. This measurement
is described by two pQCD based models~\cite{derm} when taking into account
both the imaginary and real parts of the amplitude and introducing off-diagonal
parton distributions, which leads to an enhencement of a factor $\simeq$ 5 of the 
cross section normalisation.

\section{Conclusions}

Vector meson production have been studied at HERA in a wide kinematical range 
and for different vector mesons (from the \rh\ to the $\Upsilon$ mesons).
The data show evidence for a hard behaviour of the energy
dependence for the \jpsi\ photoproduction and a similar indication is found
for \rh\ electroproduction (\qsq\ $\gsim$ 10 \gevsq).
Full helicity studies have been performed
for \rh\ and \ph\ electroproduction, showing a small but
significant violation of SCHC. The production of heavier vector mesons
($\psi'$ and $\Upsilon$) has also been observed recently at HERA.


\end{document}